\RequirePackage{ifpdf}
\ifpdf 
\documentclass[pdftex]{sigma}
\else
\documentclass{sigma}
\fi

\begin{document}

\allowdisplaybreaks

\renewcommand{\PaperNumber}{065}

\FirstPageHeading

\ShortArticleName{Emergent Supersymmetry in Warped Backgrounds}

\ArticleName{Emergent Supersymmetry in Warped Backgrounds}

\Author{Tomoaki NAGASAWA~$^{\dag^1}$, Satoshi OHYA~$^{\dag^2}$, Kazuki SAKAMOTO~$^{\dag^3\dag^4}$\\ and Makoto SAKAMOTO~$^{\dag^4}$}

\AuthorNameForHeading{T.~Nagasawa, S.~Ohya, K.~Sakamoto and M.~Sakamoto}

\Address{$^{\dag^1}$~Tomakomai National College of Technology, 443 Nishikioka, Tomakomai 059-1275, Japan}
\EmailDD{\href{mailto:nagasawa@gt.tomakomai-ct.ac.jp}{nagasawa@gt.tomakomai-ct.ac.jp}}

\Address{$^{\dag^2}$~INFN and Dipartimento di Fisica, Universit\`a di Pisa, \\
\hphantom{$^{\dag^2}$}~Largo Bruno Pontecorvo 3, 56127 Pisa, Italy}
\EmailDD{\href{mailto:satoshi.ohya@pi.infn.it}{satoshi.ohya@pi.infn.it}}

\Address{$^{\dag^3}$~Present address: Materials Research Laboratory, Kobe Steel, Ltd., Japan}

\Address{$^{\dag^4}$~Department of Physics, Kobe University, 1-1 Rokkodai, Nada, Kobe 657-8501, Japan}
\EmailDD{\href{mailto:dragon@kobe-u.ac.jp}{dragon@kobe-u.ac.jp}}

\ArticleDates{Received May 27, 2011;  Published online July 11, 2011}

\Abstract{We show that quantum mechanical supersymmetries are emerged in Kaluza--Klein spectrum of linearized gravity in several warped backgrounds as a consequence of higher-dimensional general coordinate invariance.
These emergent supersymmetries play an essential role for the spectral structure of braneworld gravity.
We show that for the case of braneworld models with two codimension-1 branes the spectral pattern is completely determined only through the supersymmetries.}


\Classification{81Q60}

\section{Introduction} \label{sec:1}

In this paper we extend our previous analysis \cite{Lim:2005rc,Lim:2007fy,Lim:2008hi} to a wider class of warped backgrounds including Randall--Sundrum model \cite{Randall:1999ee,Randall:1999vf} and Karch--Randall model \cite{Karch:2000ct}.
We will show that higher-dimensional general coordinate invariance is again translated into the quantum mecha\-ni\-cal supersymmetries in the spectrum.
The hidden supersymmetry structures we wish to illuminate in this paper are: an $N=2$ supersymmetry between graviton- and vector-modes; another $N=2$ supersymmetry between vector- and scalar-modes (with constant shift of the origin of energy); and the second-order derivative supersymmetry between graviton- and scalar-modes (with constant shift of the origin of energy).
Schematic view of this supersymmetry structure is as follows:
\begin{center}
\setlength\unitlength{1em}
\begin{picture}(33,7)(0,-3.5)
\put(4,0){(spin-0 mode)}
\put(13.5,0){(spin-1 mode)}
\put(23,0){(spin-2 mode)}

\put(11,1.0){\small $\mathbb{Q}_{(01)}^{+}$}
\put(11,-1.1){\small $\mathbb{Q}_{(01)}^{-}$}
\put(10.75,0.4){\vector(1,0){2}}
\put(12.75,0){\vector(-1,0){2}}

\put(20.5,1.0){\small $\mathbb{Q}_{(12)}^{+}$}
\put(20.5,-1.1){\small $\mathbb{Q}_{(12)}^{-}$}
\put(20.25,0.4){\vector(1,0){2}}
\put(22.25,0){\vector(-1,0){2}}

\put(15.5,2.8){\small $\mathbb{Q}_{(02)}^{+}$}
\put(7,2.2){\line(1,0){19}}
\put(7,2.2){\line(0,-1){1}}
\put(26,2.2){\vector(0,-1){1}}

\put(15.5,-3.1){\small $\mathbb{Q}_{(02)}^{-}$}
\put(7,-2){\line(1,0){19}}
\put(7,-2){\vector(0,1){1.5}}
\put(26,-2){\line(0,1){1.5}}
\end{picture}
\end{center}
where $\mathbb{Q}_{(01)}^{+}$, $\mathbb{Q}_{(12)}^{+}$, $\mathbb{Q}_{(01)}^{-}$, $\mathbb{Q}_{(12)}^{-}$ are the f\/irst-order derivative supercharges and $\mathbb{Q}_{(02)}^{+}$, $\mathbb{Q}_{(02)}^{-}$ the second-order derivative supercharges (see Section~\ref{sec:3}).
Revealing the above supersymmetry structure in several warped backgrounds without matter, we then demonstrate its impacts on spectral pattern of braneworld gravity.

The rest of this paper is organized as follows.
Section~\ref{sec:2} is devoted to a quick review for the background warped geometries we use.
In Section~\ref{sec:3} we show that quantum mechanical supersymmetries generically emerge in the Kaluza--Klein mass eigenvalue problems as a consequence of higher-dimensional general coordinate invariance.
In Section \ref{sec:4} we show that in braneworld gravity with two codimension-1 branes the spectral pattern of Kaluza--Klein modes is completely determined only through the supersymmetry structure.
We conclude in Section~\ref{sec:5}.

\section{Preliminary: background geometry} \label{sec:2}

In this paper we study linearized pure Einstein gravity on $(d+1)$-dimensional warped backgrounds described by the following metric
\begin{gather}
\mathrm{d}s^{2}
= 	G_{MN}(x, z)\mathrm{d}x^{M}\mathrm{d}x^{N}
= 	\mathrm{e}^{2A(z)}
	\left[
	g_{\mu\nu}(x)\mathrm{d}x^{\mu}\mathrm{d}x^{\nu}
	+ \mathrm{d}z^{2}
	\right], \label{eq:1}
\end{gather}
where $A(z)$ is the warp factor which turns out to play a role of superpotential (or prepotential) in analog supersymmetric quantum mechanics.
In this section we recall the background geometries given as the classical solutions to the Einstein equation without matter with respect to the metric~\eqref{eq:1}.
(Throughout of this paper the spacetime dimension $(d+1)$ is left arbitrary although in the phenomenological viewpoint we are interested in the case $d=4$.)

To begin with, let us start with the action.
The bulk Einstein--Hilbert action that describes braneworld we wish to study is
\[
S_{\text{EH}}
= 	M^{d-1}\int \mathrm{d}^{d}x\int \mathrm{d}z\sqrt{-G}
	\bigl[R(G) - d(d-1)\Lambda_{d+1}\bigr],
\]
where $M$ is the mass scale of $(d+1)$-dimensional gravity and $\Lambda_{d+1}$ is the $(d+1)$-dimensional bulk cosmological constant.
The factor $d(d-1)$ is introduced for later convenience.
$R(G)$ is the Ricci scalar curvature constructed from the background metric~$G_{MN}$.
The integration range of $z$ will be specif\/ied later.
(Our conventions for the curvature tensor, Ricci tensor etc. are summarized in Appendix~\ref{appendix:A}.)

As shown in Appendix \ref{appendix:A}, the bulk Einstein equations are reduced to the following nonlinear equations for the warp factor:
\begin{gather}
[A^{\prime}(z)]^{2} - A^{\prime\prime}(z)
= 	\Lambda_{d}, \label{eq:2}\\
A^{\prime\prime}(z)\mathrm{e}^{-2A(z)}
= 	-\Lambda_{d+1}, \label{eq:3}
\end{gather}
where prime (${}^{\prime}$) indicates the derivative with respect to $z$.
$\Lambda_{d}$ is the cosmological constant for the $d$-dimensional foliation of the bulk spacetime given by
\[
R(g) = d(d-1)\Lambda_{d},
\]
where $R(g)$ is the Ricci scalar constructed from the metric $g_{\mu\nu}(x)$.
We note that the dif\/ferential equation \eqref{eq:2} is nothing but the Riccati equation such that it can be linearized as $(-\partial_{z}^{2} + \Lambda_{d})\mathrm{e}^{-A(z)} = 0$.
Thus, according to the sign of the cosmological constants $\Lambda_{d+1}$ and $\Lambda_{d}$, we obtain the following four types of the warp factors \cite{Karch:2000ct}:
\begin{gather}
A(z)
= 	\begin{cases}
	\displaystyle
	-\log
	\left[
	\frac{\ell_{d}}{\ell_{d+1}}\sin\left(\frac{z - z_{0}}{\ell_{d}}\right)
	\right]
	& \text{for $\Lambda_{d} < 0$ and $\Lambda_{d+1} < 0$ (AdS$_{d}$/AdS$_{d+1}$)}, \\[1em]
	\displaystyle
	-\log\left(\frac{z-z_{0}}{\ell_{d+1}}\right)
	& \text{for $\Lambda_{d} = 0$ and $\Lambda_{d+1} < 0$ (M$_{d}$/AdS$_{d+1}$)}, \\[1em]
	\displaystyle
	-\log
	\left[
	\frac{\ell_{d}}{\ell_{d+1}}\sinh\left(\frac{z - z_{0}}{\ell_{d}}\right)
	\right]
	& \text{for $\Lambda_{d} > 0$ and $\Lambda_{d+1} < 0$ (dS$_{d}$/AdS$_{d+1}$)}, \\[1em]
	\displaystyle
	-\log
	\left[
	\frac{\ell_{d}}{\ell_{d+1}}\cosh\left(\frac{z - z_{0}}{\ell_{d}}\right)
	\right]
	& \text{for $\Lambda_{d} > 0$ and $\Lambda_{d+1} > 0$ (dS$_{d}$/dS$_{d+1}$)},
	\end{cases} \label{eq:4}
\end{gather}
where $z_{0}$ is the integration constant.
$\ell_{d+1}$ and $\ell_{d}$ are the curvature scale of bulk spacetime and its foliation, respectively, and given by
\[
\ell_{d+1}
:= 	\frac{1}{\sqrt{|\Lambda_{d+1}|}} \geq 0, \qquad
\ell_{d}
:= 	\frac{1}{\sqrt{|\Lambda_{d}|}} \geq 0.
\]

Now we are in a position to specify the range of coordinate~$z$.
First, without any loss of generality we can set $z_{0} = 0$ because it just corresponds to the choice of the origin.
Then, according to the conf\/iguration of codimension-1 brane(s), the range of $z$ should be chosen as follows:
\begin{itemize}\itemsep=0pt
\item Two zero-thickness branes:
\begin{gather}
z
\in 	\begin{cases}
	(z_{1}, z_{2}), \quad 0<z_{1}<z_{2}<\pi\ell_{d}, 	& \text{for AdS$_{d}$/AdS$_{d+1}$}, \\
	(z_{1}, z_{2}), \quad 0<z_{1}<z_{2}<\infty, 			& \text{for M$_{d}$/AdS$_{d+1}$ (Randall--Sundrum I \cite{Randall:1999ee})}, \\
	(z_{1}, z_{2}),\quad 0<z_{1}<z_{2}<\infty, 			& \text{for dS$_{d}$/AdS$_{d+1}$}, \\
	(z_{1}, z_{2}), \quad -\infty<z_{1}<z_{2}<\infty, 		& \text{for dS$_{d}$/dS$_{d+1}$}.
	\end{cases} \!\!\!\!\!\!\label{eq:5}
\end{gather}

\item A single zero-thickness brane:
\[
z
\in 	\begin{cases}
	(z_{1}, \pi\ell_{d}), \quad 0<z_{1}<\pi\ell_{d}, 	& \text{for AdS$_{d}$/AdS$_{d+1}$ (Karch--Randall \cite{Karch:2000ct})}, \\
	(z_{1}, \infty), \quad 0<z_{1}<\infty, 			& \text{for M$_{d}$/AdS$_{d+1}$ (Randall--Sundrum I\hspace{-.1ex}I \cite{Randall:1999vf})}, \\
	(z_{1}, \infty),\quad 0<z_{1}<\infty, 			& \text{for dS$_{d}$/AdS$_{d+1}$ (Karch--Randall \cite{Karch:2000ct})}, \\
	(z_{1}, \infty), \quad -\infty<z_{1}<\infty, 		& \text{for dS$_{d}$/dS$_{d+1}$}.
	\end{cases}
\]

\item Without zero-thickness brane:
\[
z
\in 	\begin{cases}
	(0, \pi\ell_{d}), 	& \text{for pure AdS$_{d}$/AdS$_{d+1}$}, \\
	(0, \infty), 		& \text{for pure M$_{d}$/AdS$_{d+1}$}, \\
	(0, \infty), 		& \text{for pure dS$_{d}$/AdS$_{d+1}$}, \\
	(-\infty, \infty), 	& \text{for pure dS$_{d}$/dS$_{d+1}$}.
	\end{cases}
\]
\end{itemize}
Each brane conf\/iguration has its own advantage such as a candidate to the solution of hierarchy problem \cite{Randall:1999ee} or alternative scenario to compactif\/ication as a consequence of localization of massless graviton mode at the brane position \cite{Randall:1999vf,Karch:2000ct}.
Irrespective of these brane conf\/igurations, there always exists supersymmetry structure in the spectrum of dimensional reduced theory.
For the sake of simplicity, however, in what follows we will concentrate ourselves to the case of two branes conf\/iguration  \eqref{eq:5} in order to discretize the spectrum.
The case of pure AdS$_{d}$/AdS$_{d+1}$ is brief\/ly discussed in Appendix \ref{appendix:B}.

\section{From general coordinate invariance\\ to quantum mechanical supersymmetry} \label{sec:3}

Supersymmetry structure in braneworld gravity has been already pointed out by several authors and used to analyze the Kaluza--Klein spectrum \cite{DeWolfe:1999cp,Csaki:2000fc,Miemiec:2000eq}.
However, all of these analysis are just based on one of two $N=2$ supersymmetries between graviton- and vector-modes.
Whole supersymmetry structure has not yet been uncovered.
In this section we show that quantum mechanical supersymmetries generically emerge as a consequence of $(d+1)$-dimensional general coordinate invariance.

To begin with, let us consider metric f\/luctuations $h_{MN}$ around the background metric \eqref{eq:1} as follows
\[
\mathrm{d}s^{2}
= 	\mathrm{e}^{2A(z)}
	\left[
	g_{MN}(x) + h_{MN}(x, z)
	\right]
	\mathrm{d}x^{M}\mathrm{d}x^{N}.
\]
The most useful parameterization of $h_{MN}$ is turned out to be of the form
\[
h_{MN}
= 	\begin{pmatrix}
	h_{\mu\nu} - \frac{1}{d-2}g_{\mu\nu}\phi 	& h_{\mu z} \\
	h_{z\nu} 							& \phi
	\end{pmatrix}.
\]
Under the inf\/initesimal coordinate transformation $x^{M} \mapsto \Hat{x}^{M} = x^{M} + \xi^{M}(x, z)$, the metric f\/luctuations transform, at the linearized level, as $h_{MN}(x, z) \mapsto \Hat{h}_{MN}(x, z) = h_{MN}(x, z) + \delta h_{MN}(x, z)$, where
\begin{gather}
\delta h_{\mu\nu}
=	- \nabla_{\mu}\xi_{\nu} - \nabla_{\nu}\xi_{\mu}
	- \frac{2}{d-2}g_{\mu\nu}\bigl(\partial_{z} + (d-1)A^{\prime}\bigr)\xi_{z}, \label{eq:6}\\
\delta h_{\mu z}
=	-\partial_{z}\xi_{\mu} - \nabla_{\mu}\xi_{z}, \label{eq:7}\\
\delta \phi
=	-2\bigl(\partial_{z} + A^{\prime}\bigr)\xi_{z}. \label{eq:8}
\end{gather}
Here $\nabla_{\mu}$ is the covariant derivative with respect to the background metric $g_{\mu\nu}(x)$.
As we will show below, the linearized general coordinate transformations \eqref{eq:6}--\eqref{eq:8} are translated into the supersymmetry transformations on the mode functions.

To see this, let us f\/irst suppose that the metric f\/luctuations are expanded into some complete orthogonal basis $\{f_{0}^{(n)}(z)\}$, $\{f_{1}^{(n)}(z)\}$ and $\{f_{2}^{(n)}(z)\}$, which are determined later, and written as follows
\begin{gather}
h_{\mu\nu}(x, z)
= 	\sum_{n}h_{\mu\nu}^{(n)}(x)f_{2}^{(n)}(z), \label{eq:9}\\
h_{\mu z}(x, z)
= 	\sum_{n}h_{\mu z}^{(n)}(x)f_{1}^{(n)}(z), \label{eq:10}\\
\phi(x, z)
= 	\sum_{n}\phi^{(n)}(x)f_{0}^{(n)}(z). \label{eq:11}
\end{gather}
If one wants to study braneworld models with non-compact extra dimension, contributions from the continuum spectrum must be added.
The supersymmetry structure we wish to show below is, however, independent of whether the spectrum is discrete or continuum.

Now let us move on to the analysis of supersymmetry structure between vector- and graviton-modes.
Since the covariant derivative $\nabla_{\mu}$ is blind for the coordinate $z$, the f\/irst two terms of the gauge transformation \eqref{eq:6} implies that the gauge parameter $\xi_{\mu}(x, z)$ should be expanded by the same basis to $h_{\mu\nu}$ such that it should be written as $\xi_{\mu}(x, z) = \sum_{n}\xi_{\mu}^{(n)}(x)f_{2}^{(n)}(z)$.
Then, in order to be consistent with the f\/irst term of the gauge transformation \eqref{eq:7}, $\partial_{z}f_{2}^{(n)}$ must be proportional to $f_{1}^{(n)}$.
Thus we are led to the following relation:
\begin{gather}
\mathcal{A}_{1}^{-}f_{2}^{(n)}(z)
= 	m_{n}f_{1}^{(n)}(z)
\qquad\text{with}\quad
\mathcal{A}_{1}^{-}
:= 	\partial_{z}, \label{eq:12}
\end{gather}
where $m_{n}$ is just a proportional coef\/f\/icient here.

Next, according to the second term of the gauge transformation \eqref{eq:7}, we see that the gauge parameter $\xi_{z}(x, z)$ should be expanded by the same basis to $h_{\mu z}$ such that it must be written as $\xi_{z}(x, z) = \sum_{n}\xi_{z}^{(n)}(x)f_{1}^{(n)}(z)$.
Then, according to the last term of the gauge transformation \eqref{eq:6}, we conclude that $-(\partial_{z} + (d-1)A^{\prime})f_{1}^{(n)}$ must be proportional to $f_{2}^{(n)}$:
\begin{gather}
\mathcal{A}_{1}^{+}f_{1}^{(n)}(z)
= 	m_{n}f_{2}^{(n)}(z)
\qquad\text{with}\quad
\mathcal{A}_{1}^{+}
:= 	-\bigl(\partial_{z} + (d-1)A^{\prime}\bigr), \label{eq:13}
\end{gather}
where we have used the fact that without any loss of generality we can use the same coef\/f\/icient as~\eqref{eq:12}.
These two equations lead to the following eigenvalue equations
\begin{gather}
H_{1}f_{1}^{(n)}(z)
= 	m_{n}^{2}f_{1}^{(n)}(z)
\qquad\text{with}\quad
H_{1}
:= 	\mathcal{A}_{1}^{-}\mathcal{A}_{1}^{+}
= 	-\partial_{z}\bigl(\partial_{z} + (d-1)A^{\prime}\bigr), \label{eq:14}\\
H_{2}f_{2}^{(n)}(z)
= 	m_{n}^{2}f_{2}^{(n)}(z)
\qquad\text{with}\quad
H_{2}
:= 	\mathcal{A}_{1}^{+}\mathcal{A}_{1}^{-}
= 	-\bigl(\partial_{z} + (d-1)A^{\prime}\bigr)\partial_{z}. \label{eq:15}
\end{gather}
As we will show in the next section, $\mathcal{A}_{1}^{+}$ and $\mathcal{A}_{1}^{-}	$ are hermitian conjugate to each other.
Now it is obvious that there exists an $N=2$ quantum mechanical supersymmetry structure.
Indeed, by introducing the following operators
\begin{gather*}
\mathbb{H}_{(12)}
= 	\begin{pmatrix}
	H_{1} 	& 0 \\
	0 		& H_{2}
	\end{pmatrix}
= 	\begin{pmatrix}
	\mathcal{A}_{1}^{-}\mathcal{A}_{1}^{+} 	& 0 \\
	0 								& \mathcal{A}_{1}^{+}\mathcal{A}_{1}^{-}
	\end{pmatrix}, \qquad\!
\mathbb{Q}_{(12)}^{+}
= 	\begin{pmatrix}
	0 				& 0 \\
	\mathcal{A}_{1}^{+} 	& 0
	\end{pmatrix}, \qquad\!
\mathbb{Q}_{(12)}^{-}
= 	\begin{pmatrix}
	0 & \mathcal{A}_{1}^{-} \\
	0 & 0
	\end{pmatrix},
\end{gather*}
which act on the two-component vector $(f_{1}(z), f_{2}(z))^{T}$ (where $T$ stands for transposition), we have the $N=2$ supersymmetry algebra
\begin{gather*}
\{\mathbb{Q}_{(12)}^{+}, \mathbb{Q}_{(12)}^{-}\}
= 	\mathbb{H}_{(12)}, \qquad
\{\mathbb{Q}_{(12)}^{+}, \mathbb{Q}_{(12)}^{+}\}
= 	\{\mathbb{Q}_{(12)}^{-}, \mathbb{Q}_{(12)}^{-}\}
= 	0, \\
[\mathbb{H}_{(12)}, \mathbb{Q}_{(12)}^{+}]
= 	[\mathbb{H}_{(12)}, \mathbb{Q}_{(12)}^{-}]
= 	0.
\end{gather*}

Let us proceed to f\/ind another $N=2$ supersymmetry structure between vector- and scalar-modes. The gauge transformation \eqref{eq:8} implies that $(\partial_{z} + A^{\prime})f_{1}^{(n)}$ must be proportional to $f_{0}^{(n)}$.
Thus we must have the following relation
\begin{gather}
\mathcal{A}_{0}^{-}f_{1}^{(n)}(z)
= 	\Bar{m}_{n}f_{0}^{(n)}(z)
\qquad\text{with}\quad
\mathcal{A}_{0}^{-}
:= 	\partial_{z} + A^{\prime}, \label{eq:16}
\end{gather}
where at this stage $\Bar{m}_{n}$ is introduced as a coef\/f\/icient that is independent of $m_{n}$.
A crucial step is to note the following identity of dif\/ferential operators
\begin{gather}
H_{1}
 = 	-\partial_{z}\bigl(\partial_{z} + (d-1)A^{\prime}\bigr)
 = 	- \bigl(\partial_{z} + (d-2)A^{\prime}\bigr)(\partial_{z} + A^{\prime})
	+(d-2)\Lambda_{d}, \label{eq:17}
\end{gather}
where in the last equality we have used the background Einstein equation \eqref{eq:2}.
Combining the equation \eqref{eq:17} and the eigenvalue equation \eqref{eq:14}, we get the following relation
\begin{gather}
\Bar{m}_{n}\mathcal{A}_{0}^{+}f_{0}^{(n)}(z)
= 	[m_{n}^{2} - (d-2)\Lambda_{d}]f_{1}^{(n)}(z)
\qquad\text{with}\quad
\mathcal{A}_{0}^{+}
:= 	- \bigl(\partial_{z} + (d-2)A^{\prime}\bigr). \label{eq:18}
\end{gather}
Now without any loss of generality we can set the coef\/f\/icient $\Bar{m}_{n}$ as
\[
\Bar{m}_{n}
= 	\sqrt{m_{n}^{2} - (d-2)\Lambda_{d}}.
\]
Multiplying the dif\/ferential operator $(\partial_{z} + A^{\prime})$ to \eqref{eq:18} we get the following eigenvalue equation
\begin{gather}
H_{0}f_{0}^{(n)}(z)
= 	m_{n}^{2}f_{0}^{(n)}(z)
\qquad\text{with}\quad
H_{0}
:= 	\mathcal{A}_{0}^{-}\mathcal{A}_{0}^{+} + (d-2)\Lambda_{d}. \label{eq:19}
\end{gather}
Now it is obvious that there exists another $N=2$ quantum mechanical supersymmetry structure.
Indeed, by introducing the following operators
\begin{gather*}
\mathbb{H}_{(01)}
= 	\begin{pmatrix}
	H_{0} 	& 0 \\
	0 		& H_{1}
	\end{pmatrix}
= 	\begin{pmatrix}
	\mathcal{A}_{0}^{-}\mathcal{A}_{0}^{+} 	& 0 \\
	0 								& \mathcal{A}_{0}^{+}\mathcal{A}_{0}^{-}
	\end{pmatrix}
	+ (d-2)\Lambda_{d}\mathbb{I}, \\
\mathbb{Q}_{(01)}^{+}
= 	\begin{pmatrix}
	0 				& 0 \\
	\mathcal{A}_{0}^{+} 	& 0
	\end{pmatrix}, \qquad
\mathbb{Q}_{(01)}^{-}
= 	\begin{pmatrix}
	0 	& \mathcal{A}_{0}^{-} \\
	0	& 0
	\end{pmatrix},
\end{gather*}
which act on the two-component vector $(f_{0}(z), f_{1}(z))^{T}$, we have the following algebra
\begin{gather*}
\{\mathbb{Q}_{(01)}^{+}, \mathbb{Q}_{(01)}^{-}\}
= 	\mathbb{H}_{(01)} - (d-2)\Lambda_{d}\mathbb{I}, \qquad
\{\mathbb{Q}_{(01)}^{+}, \mathbb{Q}_{(01)}^{+}\}
= 	\{\mathbb{Q}_{(01)}^{-}, \mathbb{Q}_{(01)}^{-}\}
= 	0, \\
[\mathbb{H}_{(01)}, \mathbb{Q}_{(01)}^{+}]
= 	[\mathbb{H}_{(01)}, \mathbb{Q}_{(01)}^{-}]
= 	0.
\end{gather*}
This is the $N=2$ supersymmetry algebra but the origin of energy is shifted by the constant $(d-2)\Lambda_{d}$.

Let us f\/inally study supersymmetry structure between scalar- and graviton-modes.
As was discussed in \cite{Lim:2005rc,Lim:2007fy} for the case of the Randall--Sundrum background, the symmetry that guarantees two-fold degeneracy between scalar- and graviton-modes is the second-order derivative supersymmetry, which is a nonlinear extension of ordinary $N=2$ supersymmetry discussed by \cite{Andrianov:1994aj,Andrianov:1995xt,FernandezC:1996hh,Aoyama:2001ca}.
Indeed, by introducing the operators
\begin{gather*}
\mathbb{H}_{(02)}
= 	\begin{pmatrix}
	H_{0} 	& 0 \\
	0 		& H_{2}
	\end{pmatrix}
= 	\begin{pmatrix}
	\mathcal{A}_{0}^{-}\mathcal{A}_{0}^{+} + (d-2)\Lambda_{d} 	& 0 \\
	0 									& \mathcal{A}_{1}^{+}\mathcal{A}_{1}^{-}
	\end{pmatrix}, \\
\mathbb{Q}_{(02)}^{+}
= 	\begin{pmatrix}
	0 							& 0 \\
	\mathcal{A}_{1}^{+}\mathcal{A}_{0}^{+} 	& 0
	\end{pmatrix}, \qquad
\mathbb{Q}_{(02)}^{-}
= 	\begin{pmatrix}
	0 	& \mathcal{A}_{0}^{-}\mathcal{A}_{1}^{-} \\
	0 	& 0
	\end{pmatrix},
\end{gather*}
which act on the two-component vector $(f_{0}(z), f_{2}(z))^{T}$, we have the second-order derivative supersymmetry algebra \cite{Andrianov:1994aj,Andrianov:1995xt,FernandezC:1996hh,Aoyama:2001ca}
\begin{gather}
\{\mathbb{Q}_{(02)}^{+}, \mathbb{Q}_{(02)}^{-}\}
= 	\mathbb{H}_{(02)}\bigl(\mathbb{H}_{(02)} - (d-2)\Lambda_{d}\mathbb{I}\bigr), \label{eq:20}\\
\{\mathbb{Q}_{(02)}^{+}, \mathbb{Q}_{(02)}^{+}\}
= 	\{\mathbb{Q}_{(02)}^{-}, \mathbb{Q}_{(02)}^{-}\}
= 	0, \qquad
[\mathbb{H}_{(02)}, \mathbb{Q}_{(02)}^{+}]
= 	[\mathbb{H}_{(02)}, \mathbb{Q}_{(02)}^{-}]
= 	0, \nonumber
\end{gather}
where \eqref{eq:20} follows from the intertwining relation
\begin{gather}
\mathcal{A}_{1}^{-}\mathcal{A}_{1}^{+} - \mathcal{A}_{0}^{+}\mathcal{A}_{0}^{-}
= 	(d-2)\Lambda_{d}. \label{eq:21}
\end{gather}
Note that this intertwining relation is valid only for the warp factor that satisf\/ies the background Einstein equation \eqref{eq:2}.

To summarize, we have shown that linearized general coordinate transformations reduce to the supersymmetry relations \eqref{eq:12}, \eqref{eq:13}, \eqref{eq:16}, \eqref{eq:18} for the mode functions $f_{0}^{(n)}$, $f_{1}^{(n)}$ and $f_{2}^{(n)}$.
In order for the consistency with the general coordinate invariance these mode functions must be the eigenfunctions of the Hamiltonians $H_{0}$, $H_{1}$ and $H_{2}$ and have the same eigenvalues (up to the zero-modes; see next section).
An important point to note is that the eigenvalue equations \eqref{eq:14}, \eqref{eq:15}, \eqref{eq:19} can be derived without referring equations of motion: it just follows from the general coordinate invariance.
Another important point to note is that the warp factor~$A(z)$ must be tuned to satisfy the background Einstein equation~\eqref{eq:2}, otherwise the refactorization~\eqref{eq:17} and intertwining relation~\eqref{eq:21} becomes incomplete such that the three-fold degeneracy in the spectrum will be disappeared.
Although in this paper we will not solve the Schr\"odinger equations, for the sake of completeness we summarized the corresponding Schr\"odinger Hamiltonians in Appendix~\ref{appendix:C}.

\section{Spectral pattern of two branes models} \label{sec:4}

Supersymmetry structure severely restricts the spectral structure of the model.
Indeed, as we will show below, spectral pattern of two branes model is completely determined by the supersymmetry.
To see this, we f\/irst have to specify the boundary conditions at the positions of branes.
Boundary conditions should be chosen to respect i) hermiticity of each Hamiltonian~$H_{s}$ ($s=0,1,2$) and ii) supersymmetry relations\footnote{It should be noted that in the present case the notion of hermiticity is associated to the inner product
\begin{gather}
(f_{s}, g_{s})
:= 	\int_{z_{1}}^{z_{2}} \mathrm{d}z\,\mathrm{e}^{(d-1)A(z)}
	f_{s}(z)g_{s}(z), \label{eq:22}
\end{gather}
where $f_{s}$, $g_{s}$ are square integrable functions on the interval $(z_{1}, z_{2})$.
Notice that the weight factor $\mathrm{e}^{(d-1)A(z)}$ comes from $\mathrm{e}^{(d+1)A(z)} \subset \sqrt{-G}$ times $\mathrm{e}^{-2A(z)} \subset G^{MN}R_{MN}(G)$ in the Einstein--Hilbert action.
The hermiticity requirement is then $(f_{s}, H_{s}g_{s}) = (H_{s}f_{s}, g_{s})$ for any $f_{s}$, $g_{s}$.}.
The former requirement guarantees the completeness as well as the orthogonality of the set of eigenfunctions $\{f_{s}^{(n)}(z)\}$ ($s=0,1,2$) such that it justif\/ies the mode expansions \eqref{eq:9}--\eqref{eq:11}.
It also ensures the reality of the spectrum.
The latter requirement, on the other hand, guarantees the $(d+1)$-dimensional general coordinate invariance of the theory.
As discussed in \cite{Lim:2007fy,Lim:2008hi,Nagasawa:2008an}, boundary conditions that satisfy these two requirements are uniquely determined and given by
\begin{gather}
(\mathcal{A}_{0}^{+}f_{0})(z_{i}) = 0, \label{eq:23}\\
f_{1}(z_{i}) = 0, \label{eq:24}\\
(\mathcal{A}_{1}^{-}f_{2})(z_{i}) = 0, \qquad z_{i} = z_{1}, z_{2}. \label{eq:25}
\end{gather}
Other choices of boundary conditions (say, $f_{0}(z_{i}) = 0$, $f_{1}(z_{i}) = 0$ and $f_{2}(z_{i}) = 0$) are not consistent with the supersymmetries and hence leads to the breakdown of three-fold degeneracy of the spectrum, or, equivalently, breakdown of $(d+1)$-dimensional general coordinate invariance.

\begin{figure}[t]
\centering\small
\begin{tabular}{ccccc}
\includegraphics{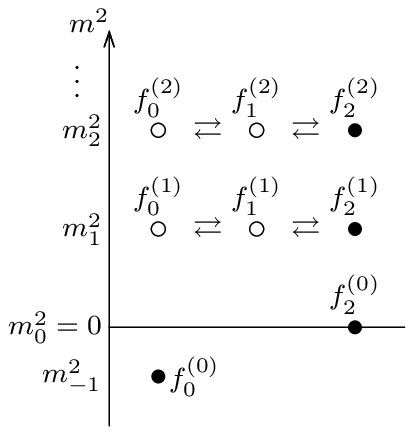} & &
\includegraphics{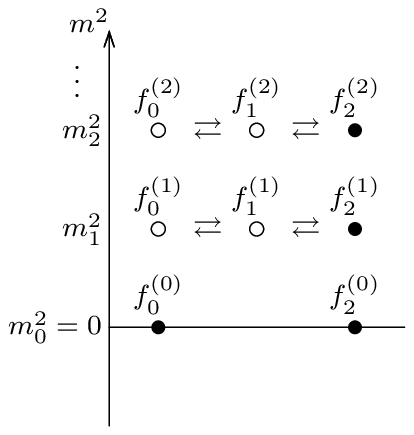} & &
\includegraphics{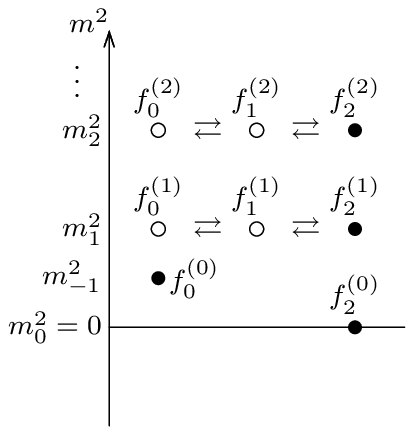} \\
(a) Case $\Lambda_{d} < 0$ (AdS$_{d}$). & &
(b) Case $\Lambda_{d} = 0$ (M$_{d}$). & &
(c) Case $\Lambda_{d} > 0$ (dS$_{d}$).
\end{tabular}
\caption{Spectral pattern of two branes models. Black dots and white circles represent physical and unphysical degrees of freedom, respectively. Up to the ground states $f_{2}^{(0)}$ and $f_{0}^{(0)}$ the spectrum exhibits three-fold degeneracy. It should be emphasized that $m_{-1}^{2} = (d-2)\Lambda_{d}$ does not directly give the radion mass; see equation \eqref{eq:32}.}
\label{fig:1}
\end{figure}

An important point to note is that with these boundary conditions the spectrum of $H_{s}$ ($s=0,1,2$) are bounded from below.
To see this, let $f_{s}$ be a normalized eigenfunction of $H_{s}$ satisfying the eigenvalue equation $H_{s}f_{s} = E_{s}f_{s}$.
Then we have
\begin{gather}
0
\leq \|\mathcal{A}_{1}^{-}f_{2}\|^{2}
= 	(\mathcal{A}_{1}^{-}f_{2}, \mathcal{A}_{1}^{-}f_{2})
= 	(f_{2}, \mathcal{A}_{1}^{+}\mathcal{A}_{1}^{-}f_{2})
= 	(f_{2}, H_{2}f_{2}) = 	E_{2}, \label{eq:26}\\
0
\leq \|\mathcal{A}_{1}^{+}f_{1}\|^{2}
 = 	(\mathcal{A}_{1}^{+}f_{1}, \mathcal{A}_{1}^{+}f_{1})
= 	(f_{1}, \mathcal{A}_{1}^{-}\mathcal{A}_{1}^{+}f_{1})
= 	(f_{1}, H_{1}f_{1}) = 	E_{1}, \\
0
\leq \|\mathcal{A}_{0}^{-}f_{1}\|^{2}
 = 	(\mathcal{A}_{0}^{-}f_{1}, \mathcal{A}_{0}^{-}f_{1})
= 	(f_{1}, \mathcal{A}_{0}^{+}\mathcal{A}_{0}^{-}f_{1})
= 	(f_{1}, [H_{1} - (d-2)\Lambda_{d}]f_{1}) \nonumber\\
\hphantom{0 \leq \|\mathcal{A}_{0}^{-}f_{1}\|^{2}}{}
= 	E_{1} - (d-2)\Lambda_{d}, \\
0
\leq \|\mathcal{A}_{0}^{+}f_{0}\|^{2}
 = 	(\mathcal{A}_{0}^{+}f_{0}, \mathcal{A}_{0}^{+}f_{0})
= 	(f_{0}, \mathcal{A}_{0}^{-}\mathcal{A}_{0}^{+}f_{0})
= 	(f_{0}, [H_{0} - (d-2)\Lambda_{d}]f_{0}) \nonumber\\
\hphantom{0\leq \|\mathcal{A}_{0}^{+}f_{0}\|^{2}}{}
= 	E_{0} - (d-2)\Lambda_{d}, \label{eq:29}
\end{gather}
where the second equalities of each line follow from the partial integration and boundary conditions \eqref{eq:23}--\eqref{eq:25}, and the third equalities the def\/initions of the Hamiltonians.
Thus we obtain the following bound of the spectrum:
\begin{gather*}
E_{2} \geq 	0, \qquad
E_{1} \geq
	\begin{cases}
	0 				& \text{for $\Lambda_{d} \leq 0$}, \\
	(d-2)\Lambda_{d} 	& \text{for $\Lambda_{d} > 0$},
	\end{cases} \qquad
E_{0} \geq 	(d-2)\Lambda_{d} =: m_{-1}^{2}.
\end{gather*}
As is evident from the expressions \eqref{eq:26}--\eqref{eq:29}, the lower bounds are saturated by the zero-modes of dif\/ferential operators $\mathcal{A}_{1}^{-}$, $\mathcal{A}_{1}^{+}$, $\mathcal{A}_{0}^{-}$ and $\mathcal{A}_{0}^{+}$, and given as the solutions to the f\/irst-order dif\/ferential equations $\mathcal{A}_{1}^{-}f_{2}^{(0)}(z) = 0$ for $E_{2} = 0$, $\mathcal{A}_{1}^{+}f_{1}^{(0)}(z) = 0$ for $E_{1} = 0$, $\mathcal{A}_{0}^{-}f_{1}^{(0)}(z) = 0$ for $E_{1} = (d-2)\Lambda_{d}$ and $\mathcal{A}_{0}^{+}f_{0}^{(0)}(z) = 0$ for $E_{0} = (d-2)\Lambda_{d}$.
These dif\/ferential equations are easily solved with the results
\begin{gather}
f_{2}^{(0)}(z)
\propto 	\mathrm{const} \hspace{3.9em} \text{for $E_{2} = 0$}, \nonumber\\
f_{1}^{(0)}(z)
\propto 	\begin{cases}
		\mathrm{e}^{-(d-1)A(z)} 	& \text{for $E_{1} = 0$ \quad ($\Lambda_{d} \leq 0$)}, \\
		\mathrm{e}^{-A(z)} 		& \text{for $E_{1} = (d-2)\Lambda_{d}$ \quad ($\Lambda_{d} > 0$)},
		\end{cases} \label{eq:30}\\
f_{0}^{(0)}(z)
\propto 	\mathrm{e}^{-(d-2)A(z)} \hspace{1.8em} \text{for $E_{0} = (d-2)\Lambda_{d}$}. \label{eq:31}
\end{gather}
Notice that in both cases $\Lambda_{d} \leq 0$ and $\Lambda_{d} > 0$ the mode function \eqref{eq:30} does not satisfy the boundary condition \eqref{eq:24} in the two branes models.
Thus the vector zero-mode must be thrown away, as it should in a respect that translational symmetry along $z$-direction is broken due to the presence of boundaries.
Since the spectrum must be discretized (because $z$-direction is compact) and further triply degenerate up to these zero-modes $\{f_{2}^{(0)}, f_{0}^{(0)}\}$, the mode expansions should become
\begin{gather*}
h_{\mu\nu}(x, z)
= 	h_{\mu\nu}^{(0)}(x)f_{2}^{(0)}(z)
	+ \sum_{n=1}^{\infty}h_{\mu\nu}^{(n)}(x)f_{2}^{(n)}(z), \\
h_{\mu z}(x, z)
= 	  0
	+ \sum_{n=1}^{\infty}h_{\mu z}^{(n)}(x)f_{1}^{(n)}(z), \\
\phi(x, z)
= 	\phi^{(0)}(x)f_{0}^{(0)}(z)
	+ \sum_{n=1}^{\infty}\phi^{(n)}(x)f_{0}^{(n)}(z),
\end{gather*}
for metric f\/luctuations, and, for gauge parameters,
\begin{gather*}
\xi_{\mu}(x, z)
= 	\xi_{\mu}^{(0)}(x)f_{2}^{(0)}(z)
	+ \sum_{n=1}^{\infty}\xi_{\mu}^{(n)}(x)f_{2}^{(n)}(z), \\
\xi_{z}(x, z)
= 	  0
	+ \sum_{n=1}^{\infty}\xi_{z}^{(n)}(x)f_{1}^{(n)}(z),
\end{gather*}
where the non-zero Kaluza--Klein modes $\{f_{2}^{(n)}, f_{1}^{(n)}, f_{0}^{(n)} \mid n \geq 1\}$ form the supersymmetry multiplets as discussed in the previous section, and share the same mass eigenvalue, $H_{s}f_{s}^{(n)} = m_{n}^{2}f_{s}^{(n)}$.
The resultant spectral pattern is depicted in Fig.~\ref{fig:1}.

\textbf{Unitary gauge.}
Now we are in a position to see the particle content of the theory and check its mass spectrum.
To this end let us go to the coordinate frame of unitary gauge.
In terms of the Kaluza--Klein modes the gauge transformations \eqref{eq:6}--\eqref{eq:8} read
\begin{gather*}
\Hat{h}_{\mu\nu}^{(n)}(x)
= 	h_{\mu\nu}^{(n)}(x)
	- \nabla_{\mu}\xi_{\nu}^{(n)}(x) - \nabla_{\nu}\xi_{\mu}^{(n)}(x)
	+ \frac{2}{d-2}g_{\mu\nu}(x)m_{n}\xi_{z}^{(n)}(x), \\
\Hat{h}_{\mu z}^{(n)}(x)
= 	h_{\mu z}^{(n)}(x)
	- m_{n}\xi_{\mu}^{(n)}(x) - \nabla_{\mu}\xi_{z}^{(n)}(x), \qquad
\Hat{\phi}^{(n)}(x)
= 	\phi^{(n)}(x) - 2\Bar{m}_{n}\xi_{z}^{(n)}(x),
\end{gather*}
for the non-zero Kaluza--Klein modes $(n \geq 1)$, and
\begin{gather*}
\Hat{h}_{\mu\nu}^{(0)}(x)
= 	h_{\mu\nu}^{(0)}(x)
	- \nabla_{\mu}\xi_{\nu}^{(0)}(x) - \nabla_{\nu}\xi_{\mu}^{(0)}(x), \qquad
\Hat{\phi}^{(0)}(x)
= 	\phi^{(0)}(x),
\end{gather*}
for the zero-modes ($n=0$).
By moving to the coordinate frame by choosing
\begin{gather*}
\xi_{z}^{(n)}(x)
= 	\frac{1}{2\Bar{m}_{n}}\phi^{(n)}(x), \qquad n\geq1, \\
\xi_{\mu}^{(n)}(x)
= 	\frac{1}{m_{n}}h_{\mu z}^{(n)}(x)
	- \frac{1}{2m_{n}\Bar{m}_{n}}\nabla_{\mu}\phi^{(n)}(x), \qquad n\geq1,
\end{gather*}
the non-zero vector- and scalar-modes are all gauged away, $\Hat{\phi}^{(n)}(x) = 0$ and $\Hat{h}_{\mu z}^{(n)}(x) = 0$, $n\geq1$.
In this coordinate frame we are left with the inf\/inite tower of massive graviton modes
\begin{gather*}
\Hat{h}_{\mu\nu}^{(n)}(x)
 = 	h_{\mu\nu}^{(n)}(x)
	- \frac{1}{m_{n}}\nabla_{\mu}h_{\nu z}^{(n)}(x)
	- \frac{1}{m_{n}}\nabla_{\nu}h_{\mu z}^{(n)}(x) \nonumber\\
\phantom{\Hat{h}_{\mu\nu}^{(n)}(x)=}{}
	+ \frac{1}{m_{n}\Bar{m}_{n}}\nabla_{\mu}\nabla_{\nu}\phi^{(n)}(x)
	+ \frac{1}{d-2}\frac{m_{n}}{\Bar{m}_{n}}g_{\mu\nu}(x)\phi^{(n)}(x), \qquad n\geq1, \nonumber
\end{gather*}
and the massless graviton mode $\Hat{h}_{\mu\nu}^{(0)}$ and the radion mode $\Hat{\phi}^{(0)}$.
These are physical degrees of freedom and turn out to satisfy the following equations of motions
\begin{gather*}
\big[\triangle_{L}^{(2)} + m_{n}^{2} - 2(d-1)\Lambda_{d}\big]\Hat{h}_{\mu\nu}^{(n)}(x)
= 	0, \qquad \nabla^{\mu}\Hat{h}_{\mu\nu}^{(n)} = g^{\mu\nu}h_{\mu\nu}^{(n)} = 0, \\
\big[\triangle_{L}^{(0)} + m_{-1}^{2} - 2(d-1)\Lambda_{d}\big]\Hat{\phi}^{(0)}(x)
= 	0,
\end{gather*}
where $\triangle_{L}^{(0,2)}$ is the Lichnerowicz operator given by
\begin{gather*}
 \bigtriangleup_{L}^{(2)}h_{\mu\nu}
 = 	- \Box_{d}h_{\mu\nu}
	+ [\nabla_{\lambda}, \nabla_{\mu}]{h^{\lambda}}_{\nu}
	+ [\nabla_{\lambda}, \nabla_{\nu}]{h_{\mu}}^{\lambda} \nonumber\\
 \phantom{\bigtriangleup_{L}^{(2)}h_{\mu\nu}}{} = 	- \Box_{d} h_{\mu\nu} - 2R_{\mu\rho\nu\sigma}(g)h^{\rho\sigma}
	+ R_{\rho\mu}(g){h^{\rho}}_{\nu} + R_{\rho\nu}(g){h^{\rho}}_{\mu},
\\
\bigtriangleup_{L}^{(0)}\phi
= 	- \Box_{d}\phi.
\end{gather*}
$\Box_{d}$ is the $d$-dimensional d'Alambertian with respect to the metric $g_{\mu\nu}(x)$.
$R_{\mu\rho\nu\sigma}(g)$ and $R_{\mu\nu}(g)$ are given in~\eqref{eq:35}, \eqref{eq:36} in Appendix~\ref{appendix:A}.
Now we are ready to read of\/f the mass of the gravitons and radion from the equations of motion.
The graviton mass is simply given by
\[
m_{\text{graviton}}^{2}
= 	m_{n}^{2}, \qquad n=0,1,2,\dots.
\]
On the other hand, the radion mass should read
\begin{gather}
m_{\text{radion}}^{2}
= 	m_{-1}^{2} - 2(d-1)\Lambda_{d}
= 	-d\Lambda_{d}, \label{eq:32}
\end{gather}
which coincides with the previous results when $\Lambda_{d}<0$ \cite{Gen:2000nu}.
Notice that for the case of de Sitter brane $\Lambda_{d} > 0$, the radion acquires negative mass squared.
Referring to the zero-mode solution~\eqref{eq:31} with the solution~\eqref{eq:4} and the inner product~\eqref{eq:22}, however, we immediately see that this de Sitter radion mode becomes non-normalizable in the limit $z_{2} \to \infty$ and hence disappears from the spectrum of single brane models as discussed by Karch and Randall~\cite{Karch:2000ct}.

\section{Conclusions} \label{sec:5}

In this paper we have studied $(d+1)$-dimensional braneworld gravity with a single extra dimension with non-vanishing bulk as well as brane cosmological constants.
Without matter, classical Einstein equation admits four distinct types of warp factors, including Randall--Sundrum and Karch--Randall models.
Irrespective of these four types of warped backgrounds, we have shown that there always exists a supersymmetry structure in the Kaluza--Klein spectrum as a consequence of $(d+1)$-dimensional general coordinate invariance.
As discussed in Section~\ref{sec:3}, we have shown that scalar- and vector-modes form $N=2$ supersymmetry multiplet, vector- and graviton-modes form another $N=2$ supersymmetry multiplet, and scalar- and graviton-modes form the second-order derivative supersymmetry multiplet.
The resultant spectrum exhibits three-fold degeneracy up to the ground states.
This supersymmetry structure is powerful enough to determine the spectral pattern of Kaluza--Klein modes.
Indeed, for the case of models with two codimension-1 branes, we have shown that the spectral pattern is controlled by supersymmetry and can be determined without referring neither equations of motion nor two-point Green functions (up to the constant shift $2(d-1)\Lambda_{d}$ for the radion mode).
What we need are only supersymmetries and boundary conditions.

\appendix

\section{Background Einstein equation} \label{appendix:A}
Let us solve the $(d+1)$-dimensional bulk Einstein equation\footnote{Our conventions are as follows:
\begin{gather}
\text{metric signature:}~
(-,+,+,\dots,+), \nonumber\\
\text{Christof\/fel symbol:}~
\Gamma^{A}_{MN}(G)
= \frac{1}{2}G^{AB}(\partial_{M}G_{BN} + \partial_{N}G_{BM} - \partial_{B}G_{MN}), \nonumber\\
\text{Curvature tensor:}~
{R^{K}}_{LMN}(G)
= \partial_{M}\Gamma^{K}_{LN}(G) - \partial_{N}\Gamma^{K}_{LM}(G)
+ \Gamma^{A}_{LN}(G)\Gamma^{K}_{AM}(G)
- \Gamma^{A}_{LM}(G)\Gamma^{K}_{NA}(G), \nonumber\\
\text{Ricci tensor:}~
R_{MN}(G)
= {R^{A}}_{MAN}(G). \nonumber
\end{gather}
}
\[
R_{MN}(G) - \frac{1}{2}G_{MN}\bigl[R(G) - d(d-1)\Lambda_{d+1}\bigr]
= 	0,
\]
with the metric ansatz
\[
G_{MN}(x, z)
= 	\mathrm{e}^{2A(z)}g_{MN}(x), \qquad
g_{MN}(x)
= 	\begin{pmatrix}
	g_{\mu\nu}(x) 	& 0 \\
	0 			& 1
	\end{pmatrix}.
\]
Regarding that $G_{MN}$ is given by the conformal transformation $g_{MN}(x) \to \mathrm{e}^{2A(z)}g_{MN}(x)$, we can easily evaluate the Ricci tensor $R_{MN}(G)$ by using its transformation law under the conformal transformation.
The result is
\begin{gather*}
R_{\mu\nu}(G)
= 	R_{\mu\nu}(g) - g_{\mu\nu}\left[A^{\prime\prime} + (d-1)(A^{\prime})^{2}\right], \qquad
R_{\mu z}(G)
= 	0, \qquad
R_{zz}(G)
= 	-dA^{\prime\prime},
\end{gather*}
where $R_{\mu\nu}(g)$ is the Ricci tensor with respect to the metric $g_{\mu\nu}(x)$.
With these expressions the Ricci scalar is given by
\begin{gather*}
R(G)
 = 	G^{\mu\nu}R_{\mu\nu}(G) + G^{zz}R_{zz}(G)
 = 	\mathrm{e}^{-2A}
	\left[
	R(g) - 2dA^{\prime\prime} - d(d-1)(A^{\prime})^{2}
	\right].
\end{gather*}
Thus the $\mu\nu$-component of bulk Einstein equation is
\begin{gather}
0
 = 	R_{\mu\nu}(g) - \frac{1}{2}g_{\mu\nu}R(g) \nonumber\\
 \phantom{0=}{}
	+ g_{\mu\nu}
	\left[
	(d-1)A^{\prime\prime} + \frac{1}{2}(d-1)(d-2)(A^{\prime})^{2}
	+ \frac{1}{2}d(d-1)\Lambda_{d+1}\mathrm{e}^{2A}
	\right], \label{eq:33}
\end{gather}
while the $zz$-component is
\begin{gather}
0
= 	-\frac{1}{2}R(g) + \frac{1}{2}d(d-1)(A^{\prime})^{2}
	+ \frac{1}{2}d(d-1)\Lambda_{d+1}\mathrm{e}^{2A}. \label{eq:34}
\end{gather}
Note that the $\mu z$-component is trivial and does not lead to any constraint.
Subtracting \eqref{eq:33} by $g_{\mu\nu} \times \eqref{eq:34}$ we get
\[
R_{\mu\nu}(g)
= 	(d-1)g_{\mu\nu}[(A^{\prime})^{2} - A^{\prime\prime}].
\]
By contracting this expression with respect to $g_{\mu\nu}$ and substituting the result into the equation~\eqref{eq:34} we get
\begin{gather*}
R(g)
= 	d(d-1)[(A^{\prime})^{2} - A^{\prime\prime}], \qquad
0
= 	A^{\prime\prime} + \Lambda_{d+1}\mathrm{e}^{2A}.
\end{gather*}
Since the warp factor $A(z)$ is a function of $z$ while the scalar curvature $R(g)$ a function of $x^{\mu}$, there is no nontrivial solution to the Einstein equation except for the constant curvature case $R(g) = \mathrm{const}$.
Thus, by setting $R(g) = d(d-1)\Lambda_{d}$, we obtain the announced equations~\eqref{eq:2}, \eqref{eq:3}.
We note that with these background metric the following identities hold
\begin{gather}
R_{\mu\rho\nu\sigma}(g)
= 	\Lambda_{d}(g_{\mu\nu}g_{\rho\sigma} - g_{\mu\sigma}g_{\nu\rho}), \label{eq:35}\\
R_{\rho\sigma}(g)
= 	{R^{\mu}}_{\rho\mu\sigma}(g)
= 	(d-1)\Lambda_{d} g_{\mu\nu}. \label{eq:36}
\end{gather}

\section[Shape invariance method and graviton mass spectrum for pure AdS$_{d}$/AdS$_{d+1}$]{Shape invariance method and graviton mass spectrum\\ for pure $\boldsymbol{{\rm AdS}_{d}/{\rm AdS}_{d+1}}$} \label{appendix:B}

\begin{figure}[t]
\centering
\includegraphics{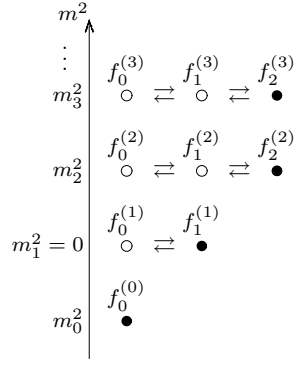}
\caption{Spectral pattern of pure AdS$_{d}$/AdS$_{d+1}$. Particle contents are: one massive scalar ($f_{0}^{(0)}$), one massive vector ($f_{1}^{(1)}$), and an inf\/inite tower of massive gravitons ($\{f_{2}^{(n)} \mid n=2,3,4,\dots\}$). The spectrum of Hamiltonian $H_{0}$ is given by $m_{n}^{2} = (n-1)(n+d-2)|\Lambda_{d}|$, $n=0,1,2,\dots$.}
\label{fig:2}
\end{figure}

Similar analysis presented in Section \ref{sec:3} leads to the following hierarchy of Hamiltonians
\begin{alignat}{7}
\text{spin-$0$ mode}:~
&H_{0} = \mathcal{A}_{0}^{-}\mathcal{A}_{0}^{+} + \varepsilon_{0}
&
&
&
&
&
&\nonumber\\
\text{spin-$1$ mode}:~
&H_{1} = \mathcal{A}_{0}^{+}\mathcal{A}_{0}^{-} + \varepsilon_{0}~
&
&= \mathcal{A}_{1}^{-}\mathcal{A}_{1}^{+} + \varepsilon_{1}
&
&
&
&\nonumber\\
\text{spin-$2$ mode}:~
&H_{2} =
&
&= \mathcal{A}_{1}^{+}\mathcal{A}_{1}^{-} + \varepsilon_{1}~
&= \mathcal{A}_{2}^{-}\mathcal{A}_{2}^{+} + \varepsilon_{2}
&
&\nonumber\\
\text{spin-$3$ mode}:~
&H_{3} =
&
&
&= \mathcal{A}_{2}^{+}\mathcal{A}_{2}^{-} + \varepsilon_{2}
&= \cdots
&\nonumber\\
&~~\vdots
&
&
&
&\,\,\,\vdots
& \nonumber
\end{alignat}
For general $s$, the f\/irst-order dif\/ferential operators $\mathcal{A}_{s}^{+}$ and $\mathcal{A}_{s}^{-}$ are given by
\begin{gather*}
\mathcal{A}_{s}^{+}
= 	- \partial_{z} - (s + d - 2)A^{\prime}(z), \qquad
\mathcal{A}_{s}^{-}
= 	+ \partial_{z} - (s-1)A^{\prime}(z),
\end{gather*}
which satisf\/ies the intertwining relation
\[
\mathcal{A}_{s}^{-}\mathcal{A}_{s}^{+} - \mathcal{A}_{s-1}^{+}\mathcal{A}_{s-1}^{-}
= 	2\Bar{s}\Lambda_{d},
\]
where
\[
\Bar{s} := 	s + \frac{d-4}{2}.
\]
The constant shift $\varepsilon_{s}$ is given by
\[
\varepsilon_{s}
= 	- (s-1)(s+d-2)\Lambda_{d}
= 	\left[\frac{(d-1)^{2}}{4} - \left(\Bar{s} + \frac{1}{2}\right)^{2}\right]\Lambda_{d}.
\]
Notice that when there is no codimension-1 branes, standard shape invariance method is applicable (for shape invariance method, see for example \cite{Cooper:1994eh}).
Thus, for pure AdS$_{d}$/AdS$_{d+1}$, the graviton mass spectrum can be obtained without solving the equation of motion and given by
\[
m_{\text{graviton}}^{2}
= 	m_{n}^{2}
:= 	(n-1)(n+d-2)|\Lambda_{d}|, \qquad n=2,3,4,\dots,
\]
which coincides with the group theoretical results when $d=4$ \cite{Karch:2000ct}.
The resultant spectral pattern becomes as shown in Fig.~\ref{fig:2}.

\section{Analog supersymmetric quantum mechanics} \label{appendix:C}
Under the following similarity transformation
\[
h_{MN}
\mapsto
\Bar{h}_{MN}
= 	\mathrm{e}^{\frac{d-1}{2}A(z)}h_{MN},
\]
which eliminates the weight factor $\mathrm{e}^{(d-1)A(z)}$ in the inner product \eqref{eq:22}, the Hamiltonian is transformed as $H_{s} \mapsto \Bar{H}_{s} = \mathrm{e}^{\frac{d-1}{2}A(z)}H_{s}\mathrm{e}^{-\frac{d-1}{2}A(z)}$, or,  explicitly,
\[
\Bar{H}_{s}
= 	\Bar{\mathcal{A}}_{s}^{-}\Bar{\mathcal{A}}_{s}^{+} + \varepsilon_{s}
= 	\Bar{\mathcal{A}}_{s-1}^{+}\Bar{\mathcal{A}}_{s-1}^{-} + \varepsilon_{s-1},
\]
where $\Bar{\mathcal{A}}_{s}^{+}$ and $\Bar{\mathcal{A}}_{s}^{-}$ are the similarity transformed f\/irst-order dif\/ferential operators given by
\begin{gather*}
\Bar{\mathcal{A}}_{s}^{+}
= 	\mathrm{e}^{\frac{d-1}{2}A(z)}\mathcal{A}_{s}^{+}\mathrm{e}^{-\frac{d-1}{2}A(z)}
= 	+\partial_{z} + \left(\Bar{s} + \frac{1}{2}\right)A^{\prime}(z), \\
\Bar{\mathcal{A}}_{s}^{-}
= 	\mathrm{e}^{\frac{d-1}{2}A(z)}\mathcal{A}_{s}^{-}\mathrm{e}^{-\frac{d-1}{2}A(z)}
= 	-\partial_{z} + \left(\Bar{s} + \frac{1}{2}\right)A^{\prime}(z).
\end{gather*}
With this similarity transformation the f\/irst-order derivative terms disappear from the Hamiltonians.
Indeed, by substituting the background solution \eqref{eq:4}, the similarity transformed Hamiltonian reads
\[
\Bar{H}_{s}
= 	-\partial_{z}^{2} + V_{s}(z),
\]
where the potential is given by
\[
V_{s}
= 	(\Bar{s}^{2} - 1/4)A^{\prime\prime}(z)
	+ \frac{(d-1)^{2}}{4}\Lambda_{d},
\]
or, more explicitly,
\begin{align}
V_{s}(z)
&= 	\begin{cases}
	\displaystyle
	\frac{1}{\ell_{d}^{2}}
	\frac{\Bar{s}^{2} - 1/4}{\sin^{2}(z/\ell_{d})}
	- \frac{(d-1)^{2}}{4}\frac{1}{\ell_{d}^{2}}
	& \text{for AdS$_{d}$/AdS$_{d+1}$}, \vspace{2mm}\\
	\displaystyle
	\frac{\Bar{s}^{2} - 1/4}{z^{2}}
	& \text{for M$_{d}$/AdS$_{d+1}$}, \vspace{2mm}\\
	\displaystyle
	\frac{1}{\ell_{d}^{2}}
	\frac{\Bar{s}^{2} - 1/4}{\sinh^{2}(z/\ell_{d})}
	+ \frac{(d-1)^{2}}{4}\frac{1}{\ell_{d}^{2}}
	& \text{for dS$_{d}$/AdS$_{d+1}$}, \vspace{2mm}\\
	\displaystyle
	-\frac{1}{\ell_{d}^{2}}
	\frac{\Bar{s}^{2} - 1/4}{\cosh^{2}(z/\ell_{d})}
	+ \frac{(d-1)^{2}}{4}\frac{1}{\ell_{d}^{2}}
	& \text{for dS$_{d}$/dS$_{d+1}$}.
	\end{cases} \label{eq:37}
\end{align}
Thus the spectral problem of our braneworld gravity just reduces to the problem of supersymmetric quantum mechanics with the trigonometric P\"oschl--Teller potential, inverse square potential, and hyperbolic P\"oschl--Teller potential of $\sinh$ and $\cosh$ types.
Notice that the constant term in~\eqref{eq:37} is nothing but the Breitenlohner--Freedman (BF) bound in AdS$_{d}$ \cite{Breitenlohner:1982bm}:
\[
m_{\text{BF}}^{2}
= 	- \frac{(d-1)^{2}}{4}\frac{1}{\ell_{d}^{2}}.
\]

\subsection*{Acknowledgements}
This work is supported in part by a Grant-in-Aid for Scientif\/ic Research (No. 22540281 (M.S.)) from the Japanese Ministry of Education, Science, Sports and Culture.

\pdfbookmark[1]{References}{ref}
\LastPageEnding

\end{document}